\documentclass[aip,apl,groupedaddress,superscriptaddress,showkeys,twocolumn,a4paper]{revtex4}
\usepackage{graphicx,soul}
\usepackage[left=1.7cm,top=1.7cm,right=1.7cm,bottom=1.7cm]{geometry}
\begin{document}
\parskip=0cm
\title{Linear non-hysteretic gating of a very high density 2DEG in an undoped metal-semiconductor-metal sandwich structure}

\author{K. Das Gupta}\email{kd241@cam.ac.uk,   kdasgupta@phy.iitb.ac.in}
\affiliation{Cavendish Laboratory, J.J. Thomson Avenue, Cambridge CB3 0HE, UK.}
\affiliation{Indian Institute of Technology Bombay, Mumbai 400 076, India}
\author{A. F. Croxall} \author{W.Y. Mak}
\author{H. E. Beere}, \author{C. A. Nicoll},
\author{I. Farrer} \author{F. Sfigakis} \author{D. A. Ritchie}
\affiliation{Cavendish Laboratory, J.J. Thomson Avenue, Cambridge CB3 0HE, UK.}

\begin{abstract}

Modulation doped GaAs-AlGaAs quantum well based structures are usually used to
achieve very high mobility 2-dimensional electron (or hole) gases. Usually high
mobilities ($>10^{7}{\rm{cm}^{2}\rm{V}^{-1}\rm{s}^{-1}}$) are achieved at high
densities.  A loss of linear gateability is often associated with the highest
mobilites, on account of a some residual hopping or parallel conduction in the
doped regions. We have developed a method of using fully undoped GaAs-AlGaAs
quantum wells, where densities $\approx{6\times10^{11}\rm{cm}^{-2}}$ can be
achieved while maintaining fully linear and non-hysteretic gateability. We use
these devices to understand the possible mobility limiting mechanisms at very
high densities.

\end{abstract}

\pacs{73.40.Kp, 73.20.Mf}
\keywords{undoped quantum well, double-side
processing, etch-stop layer, backgate}

\maketitle

The ability to tune the carrier density of a 2-dimensional electronic system
(2DES) over large ranges with a linear and non-hysteretic gate is one of the
most  desirable and generic aspects in experiments that involve a 2DES.
Fundamental aspects of a  2DES, like the ratio of Coulomb and kinetic energy,
screening, relative importance of various scattering mechanisms are all
functions of the carrier density. The low density end
($\sim{10^{9}\rm{cm}^{-2}}$ and lower) is of great interest because the very
dilute 2DES is a strongly interacting system\cite{Tsui2007PRL}, where the Coulomb
interaction energy outweighs the kinetic energy. One the other hand the very
high density end ($10^{11}-10^{12}\rm{cm}^{-2}$) is of interest because the
highest electron mobilities \cite{UmanskyJCG2009}can be achieved at these
densities. Qualitatively, this happens because the effect of ionized impurity
scattering diminishes as $k_{F}$ (the Fermi wavevector) becomes larger compared
to the Fourier components of the impurity potential (${\sim}e^{-qd}/q$, where q
is the scattering wavevector and $d$ is the distance of the ionized impurity
from the plane of the 2DES ). Study of several other phenomena like non-parabolic
effects and anti-crossing of hole bands\cite{WinklerBook} , mobility limiting
effect of interface roughness \cite{Mak2010APL}, study of novel Fractional
Quantum Hall (FQHE) states \cite{Nuebler2010PRB,DasSarma2010PRB}also require
single-subband, parallel-conduction free, linearly gateable, non-hysteretic
2DES in the density
range ($10^{11}-10^{12}\rm{cm}^{-2}$).\\

\begin{figure}
\includegraphics[width=0.49\textwidth,clip=]{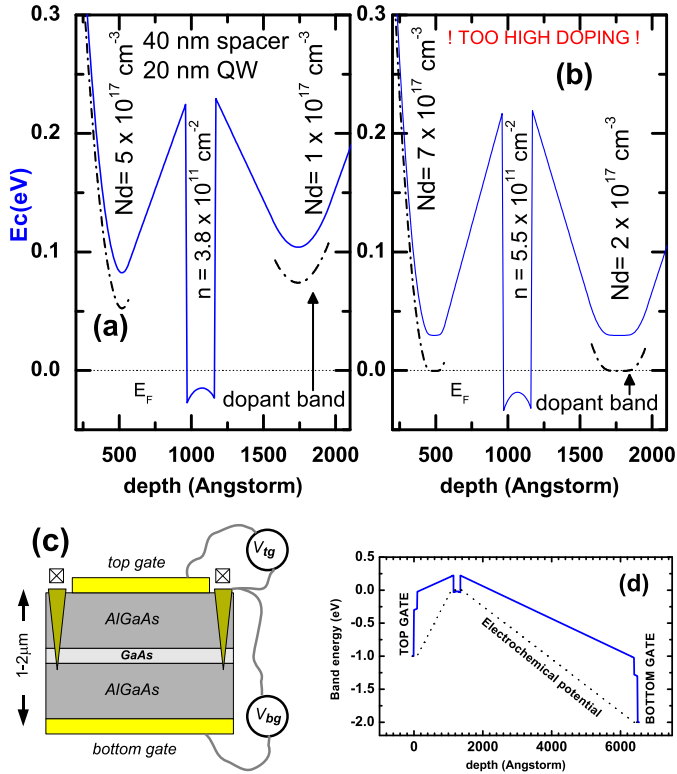}
\caption{\label{basic-idea-figure} (Colour online)(a) \& (b) Example of a  self
consistent band structure around a typical GaAs-AlGaAs quantum well. Notice how
the dopant band drops and touches the electrochemical potential at high doping
levels.(c) The basic idea behind the fully undoped metal-semiconductor-metal
sandwich structure, that can be kept free of parallel conduction at high
densities (d) The expected variation of the electrochemical potential across
the sample. The numbers are approximate.}
\end{figure}

 The advent of the quantum well structure with modulation doping
 \cite{Dingle1978APL}
 and an undoped spacer allowed higher densities and mobilities to be reached compared to what
was possible with a heterostructure. Such structures have been the workhorse
for 2DES based devices for last 30 years. But a limitation of this scheme
becomes apparent at high densities (Fig. \ref{basic-idea-figure} (a)\&(b)). As
the (as grown) carrier density is increased by increasing the doping
concentration the slope of the conduction band (CB) just outside the well must
also increase. This is necessary to satisfy electrostatics, because the flux of
the electric field (slope of the conduction band) over a box that encloses the
quantum well must equal the charge contained within. But the sharper slope of
the conduction band forces the impurity band (${\sim}30$ meV below CB in
Al$_{0.3}$Ga$_{0.7}$As) to come closer to the electrochemical potential. Thus
high doping also makes devices more prone to unwanted hopping/parallel
conduction.Variants of the modulation doping scheme like
short-period-superlattice (SPSL) \cite{UmanskyJCG2009} doping cannot circumvent
this basic problem. In fact it has been acknowledged in recent literature that
the highest mobility 2DES also suffers from some amount of hysteresis and is
often not linearly gateable owing to slow charge transfer/relaxation in the
dopant layers\cite{RosslerNJP2010}. The significance of linear gateability at
high densities for understanding the ($\nu=5/2$) FQHE state have also been
highlighted recently\cite{Nuebler2010PRB,DasSarma2010PRB}. Indeed the presence
of small amounts of parallel conducting channels would cause the Hall plateaus
to loose  exact quantization and the zeros of the Shubnikov de-Haas
oscillations to loose their sharp definition. We shall show in this paper that
gateability of the 2DES is also crucial for positioning the envelope of the
carrier wavefunction in the quantum well for obtaining the maximum mobility.\\

We describe the growth, fabrication and initial measurements on devices where
linear gating is demonstrated from $<4{\times}10^{10}\rm{cm}^{-2}$ to
${\sim}6{\times}10^{11}\rm{cm}^{-2}$ in a 20nm wide (GaAs-AlGaAs) quantum well.
The highest electron mobility we achieve is
$\mu_{e}{\approx}9{\times}10^{6}\rm{cm}^{2}\rm{V}^{-1}s^{-1}$ at T=1.5K.
We present some data indicative of the mobility limiting scattering mechanism
at the highest density. Finally we discuss, how the range of the densities can
be increased further and the possible iterative
improvement of growth conditions to improve the mobilities. \\

\begin{figure}
\begin{center}
\includegraphics[width=0.48\textwidth,clip]{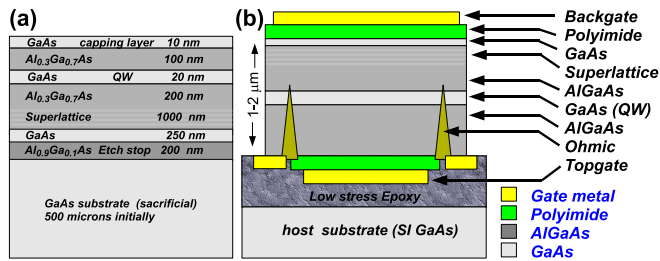}
\caption{\label{wafer-and-device-schematic-figure} (a) Layering of the wafer
used to make the devices. (b) Schematic of the device. Note that the device is
turned upside down after finishing the topside processing.}
\end{center}
\end{figure}

Fig. \ref{basic-idea-figure}{(c) \& (d)} shows the basic idea behind our
method. If a very thin (1-2$\mu$m), completely undoped AlGaAs-GaAs-AlGaAs can
be sandwiched between two layers of metals, then the backgate and topgate bias
on these plates ($V_{BG}$ and $V_{TG}$ w.r.t. the ohmics which connect to the
quantum well) can be used to attract carriers into the well. There is no
intentional dopant layer anywhere. The electrochemical potential itself goes
down as one moves out from the quantum well, because the positive voltage bias
on the gates are set to attract electrons (positive bias lowers the
electrochemical potential). The combination of these two ensure that there is
no place for parallel conduction to develop. Also the relative bias on the two
gates can be adjusted to tune the shape and tilt of the wavefunction in the
quantum well. Two practical considerations are however needed at this point.
First there must be ohmic contacts going into the quantum well. Second, the
1-2$\mu$m thick sandwiched structure cannot be self-supporting. Thus the
fabrication method has to ensure alignment of the topside and bottom side
features, as well as stress-free embedding of this structure in a suitably
rigid base. Our fabrication method achieves these . The packaged devices showed
no cracking   after several thermal cycles from room temperature to 1.5K. The
densities and mobilities obtained from
measurements in two different cryostats agreed within ${\sim}10\%$\\

The wafer used in this study was grown on a 500$\mu$m [100] GaAs substrate as
shown in Fig. \ref{wafer-and-device-schematic-figure}a. The quantum well was
located approximately 100nm below the surface and the etch stop
(Al$_{0.9}$Ga$_{0.1}$As) was approximately 1500nm below the surface. The
topside processing consisted of four main steps. A Hall bar shaped mesa was
etched (150-200nm) and ohmics were lithographically defined. AuGeNi contacts
were annealed at 450$^{o}$C for 180sec in a reducing atmosphere
(N$_{2}$/H$_{2}$) after liftoff. A layer of polyimide (HD 4104, HD
microsystems) was then spin coated on the sample. The polyimide layer was
400-500nm thick after curing at 250$^{o}$C. A metal topgate (Ti/Au) was
patterned and deposited on top of the polyimide layer. After this the sample
was embedded topside down on a host (GaAs wafer) with a thin layer of epoxy
.  The GaAs substrate was then removed  from the back using a
combination of abrasive mechanical polishing and selective etching (in
Citric-acid+H$_{2}$O$_{2}$) etch to expose the Al$_{0.9}$Ga$_{0.1}$As etch
stop. The etch stop was then  removed using Hydrofluoric acid (HF). The HF
removes the etch stop due to its high Al concentration but does not attack the
GaAs layer below. At this stage a smooth mirror finish is obtained. A  careful
inspection is done for any cracks and deformations. The sample is then coated
with another layer of polyimide and the backgate is deposited on top of the
cured polyimide. During this last stage the bond pads to the sample are also
defined, which are then used to connect the ohmics and the gates to a leadless
chip carrier using an ultrasonic wire bonder. A similar use (with some
differences) of epoxy embedding and etch-stop layer was introduced  by
Weckworth {\it et al}\cite{Weckworth1996SLM} for fabricating backgates on a
bilayer 2DES.

\begin{figure}
\includegraphics[width=0.49\textwidth,clip=]{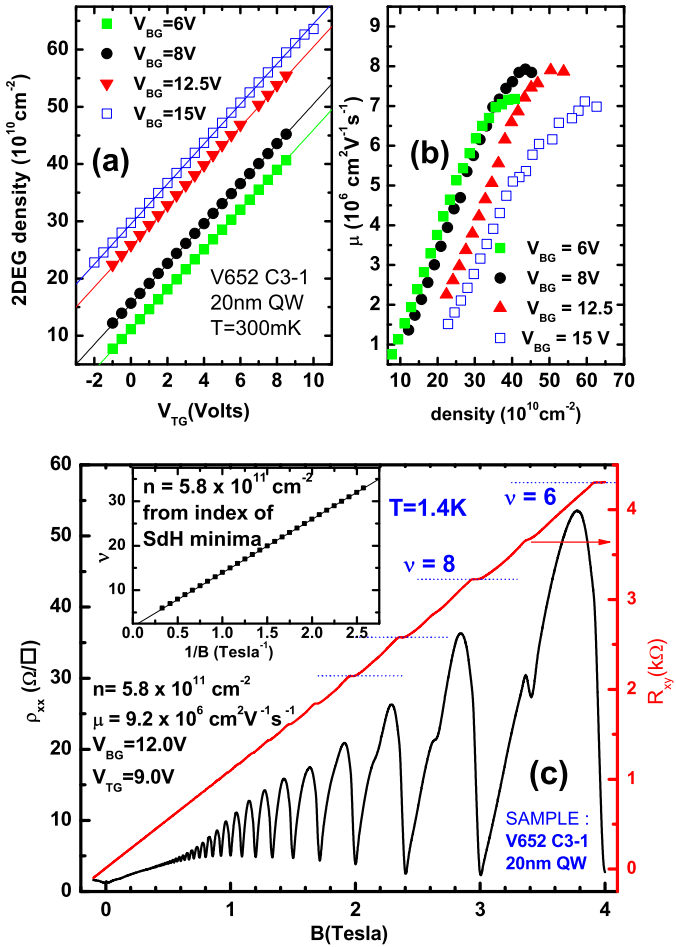}
\caption{\label{linear-gating-hall-and-sdh} (Colour online)(a) Linear
gateability of the 2-dimensional electron gas (2DEG)till
$6{\times}10^{11}\rm{cm}^{-2}$ (b) the mobility of the 2DEG as a function of
density for different backgate voltages. (c) Agreement of the slope of the Hall
resistance and the period of oscillation of the Shubnikov de-Haas minima, shows
that the 2DEG is indeed in a single subband regime.}
\end{figure}

Fig \ref{linear-gating-hall-and-sdh}a shows that combination of the topgate
voltage ($V_{TG}$) and bottom gate voltage ($V_{BG}$) induces a  2-dimensional
electron gas (2DEG) in the quantum well. The proof that the channel indeed
forms in the quantum well and not somewhere else is provided by the clean
Quantum Hall traces. It  is impossible to have a high mobility (close to
$10^{7}{\rm cm^{2}V^{-1}s^{-1}}$ ) single subband 2DEG anywhere else ({\it
e.g.} the superlattice). The single subband nature of the 2DEG is established
by the complete agreement (better than 1 part in 100) of the densities
calculated from the slope of the Hall voltage and the period of the
oscillations. From the slope of the density of the 2DEG ($n$)vs $V_{TG}$ and
$V_{BG}$ we  can calculate the capacitance of the gates to the 2DEG. This
agrees well with the calculated values obtained using the known thickness of
the semiconductor and polyimide layers.
\begin{equation}
n = \frac{C_{TG}}{e}V_{TG} + \frac{C_{BG}}{e}(V_{BG} -V_{0})
\label{n-vs-vtg-vbg-eqn}
\end{equation}
where $C_{TG}$ and $C_{BG}$ denote the topgate and backgate capacitances. We
associate a threshold voltage ($V_{0}$) with the backgate because the specific
design of our device uses the backgate to activate the ohmics.
$V_{0}\sim4{\rm{V}}$ in our devices.
\\

Since both $V_{TG}$ and $V_{BG}$ may be used to tune the carrier density, it is
possible to obtain the same density for a number of combinations of the two
voltages. In these different combinations $n$ is  same but the shape of the
wavefunction is different. Fig \ref{linear-gating-hall-and-sdh}b shows that the
mobility of the electron gas can vary by nearly 50\% depending on the choice of
the two gate voltages for the same $n$. For example if we examine the four
traces in Fig \ref{linear-gating-hall-and-sdh}a, we find that the density of
$n=4{\times}10^{11}{\rm{cm}}^{-2}$ may be obtained by setting $V_{BG}$=6V and
$V_{TG}$=8.25V, (green filled square) or $V_{BG}$=15V and $V_{TG}$=3.0V (blue
empty square). Fig \ref{linear-gating-hall-and-sdh}b shows that the mobility in
the first case $7.2{\times}10^{6}\rm{cm}^{2}\rm{V}^{-1}\rm{s}^{-1}$ but in the
second case it is only $5.0{\times}10^{6}\rm{cm}^{2}\rm{V}^{-1}\rm{s}^{-1}$.
Indeed the large difference in the four traces in Fig.
\ref{linear-gating-hall-and-sdh}b indicates that in the highest mobility regime
the ionized background is not necessarily the dominant factor determining the
mobility. In a quantum well, we necessarily have two GaAs-AlGaAs interfaces.
Inevitably the  first of the interfaces (in order of growth) is "inverted"
(GaAs is grown on top of AlGaAs) and is thought to be have more interface
roughness than the other interface in which AlGaAs is grown on GaAs. The
relative proximity of the wavefunction to these interfaces will determine the
amount of interface roughness scattering experienced by the electrons. In our
devices a larger $V_{BG}$ leads to lower mobilities because the wavefunction is
then tilted more towards the inverted interface.  In both cases the electrons
see a similar ionized background resulting from the unintentional impurities
incorporated during MBE growth remains the same\cite{VChamberBackgr}. The small
change in the form factor of the wavefunction resulting from the change in
tilt, cannot account for the large change in screening (or the dielectric
function) that would be required to account for a large change in mobility.
Coulomb scattering arising from ionized impurities cannot account for this
change, leaving the roughness of the
interface as the only possible source of the observed change in mobility.\\

Undoped heterostructures  have been known to be particularly useful for
maintaining high mobility at low densities
\cite{Tsui2007PRL,Kane1993APL,Harrell1999APL,Sarkozy2009APL}, making very
shallow gateable 2DEGs \cite{Mak2010APL}. In this paper we have shown that the
field-effect mechanism of pulling carriers from the ohmics into the conducting
channel, can be useful in reaching very high densities as well. Our method
applies equally well for creating an electron or a hole type channel. Our
method requires the MBE chamber used for growth  to be optimized for the lowest
possible unintentional background and interface roughness once - and not
separately for n-type/p-type dopants (which is essential if modulation doping,
or any of its variants, is used). The experimenter can decide whether to
fabricate n-type or p-type ohmic contacts. It is clearly possible to make these
devices fully ambipolar by fabricating both n-type and p-type ohmics on the
same Hall-bar. We anticipate that gateable higher densities could also possibly
be used to increase the energy gaps of fragile FQHE states by forcing the same
filling factor to occur at higher magnetic fields.

\bibliographystyle{apsrev4-1}

\end{document}